\documentclass[aip, apl, reprint]{revtex4-1}

\pdfoutput=1 
\usepackage{graphicx}
\usepackage{color,amsmath}
\usepackage[normalem]{ulem}

\usepackage[colorlinks,linkcolor=blue,citecolor=blue,urlcolor=blue]{hyperref}
\usepackage{siunitx}
\usepackage{comment}
\usepackage{layouts}

\begin{document}
\title{Edgeless and Purely Gate-Defined Nanostructures in InAs Quantum Wells}

\author{Christopher Mittag}
\email{mittag@phys.ethz.ch}
\affiliation{Solid State Physics Laboratory, Department of Physics, ETH Zurich, 8093 Zurich, Switzerland}

\author{Matija Karalic}
\affiliation{Solid State Physics Laboratory, Department of Physics, ETH Zurich, 8093 Zurich, Switzerland}

\author{Zijin Lei}
\affiliation{Solid State Physics Laboratory, Department of Physics, ETH Zurich, 8093 Zurich, Switzerland}

\author{Thomas Tschirky}
\affiliation{Solid State Physics Laboratory, Department of Physics, ETH Zurich, 8093 Zurich, Switzerland}

\author{Werner Wegscheider}
\affiliation{Solid State Physics Laboratory, Department of Physics, ETH Zurich, 8093 Zurich, Switzerland}

\author{Thomas Ihn}
\affiliation{Solid State Physics Laboratory, Department of Physics, ETH Zurich, 8093 Zurich, Switzerland}

\author{Klaus Ensslin}
\affiliation{Solid State Physics Laboratory, Department of Physics, ETH Zurich, 8093 Zurich, Switzerland}

\date{\today}

\begin{abstract}
Nanostructures in InAs quantum wells have so far remained outside of the scope of traditional microfabrication techniques based on etching. This is due to parasitic parallel conduction arising from charge carrier accumulation at the physical edges of samples. Here we present a technique which enables the realization of quantum point contacts and quantum dots in two-dimensional electron gases of InAs purely by electrostatic gating. Multiple layers of top gates separated by dielectric layers are employed. Full quantum point contact pinch-off and measurements of Coulomb-blockade diamonds of quantum dots are demonstrated. 
\end{abstract}

\maketitle
InAs is a semiconductor material of strong spin-orbit interaction, low effective mass and large g-factor, for which it has gained recent interest. It arises, for instance, from proposals for investigating topological phenomena. One-dimensional InAs nanowires combined with superconductors are expected to be a host system for Majorana physics\,\cite{kitaev_unpaired_2001, oreg_helical_2010, lutchyn_majorana_2010, das_zero-bias_2012, mourik_signatures_2012}. Composite quantum wells were proposed to show the quantum spin Hall effect\,\cite{kane_quantum_2005, liu_quantum_2008, knez_evidence_2011, qu_electric_2015, mueller_nonlocal_2015, suzuki_edge_2013}. For potential future applications of topological quantum computation which require upscaling to a large number of qubits, it would be advantageous to start from a 2-dimensional structure\,\cite{suominen_zero-energy_2017, vaitiekenas_selective_2018, lee_selective-area_2018}, which would severely simplify integration. Therefore, developing functional nanostructures made from InAs two-dimensional electron gases is a natural starting point. Control over Coulomb islands or nanostructures is paramount for these kinds of experiments.

Yet, up to date, convincing nanostructures in InAs two-dimensional electron gases have not been realized with the exception of a few attempts with trench-etched quantum point contacts\,\cite{koester_length_1996, debray_all-electric_2009, matsuo_magnetic_2017}. We believe that the reason lies in the trivial edge conduction recently reported in InAs\,\cite{de_vries_$h/e$_2018, mittag_passivation_2017} and InAs/GaSb quantum wells\,\cite{nichele_edge_2016, nguyen_decoupling_2016, mueller_edge_2017}.
However, in InAs nanowires this problem seems not to occur and exquisit control over single and multi-dot systems has been reported\,\cite{fasth_direct_2007}.
For InAs, the Fermi level is pinned in the conduction band at the surface\,\cite{noguchi_intrinsic_1991, olsson_charge_1996}. The origin of this Fermi level pinning is not clear yet, but possible reasons are discussed in Refs.\,\onlinecite{de_vries_$h/e$_2018}\,and\,\onlinecite{nichele_edge_2016}.

\begin{figure}[]
	\includegraphics[width=\columnwidth]{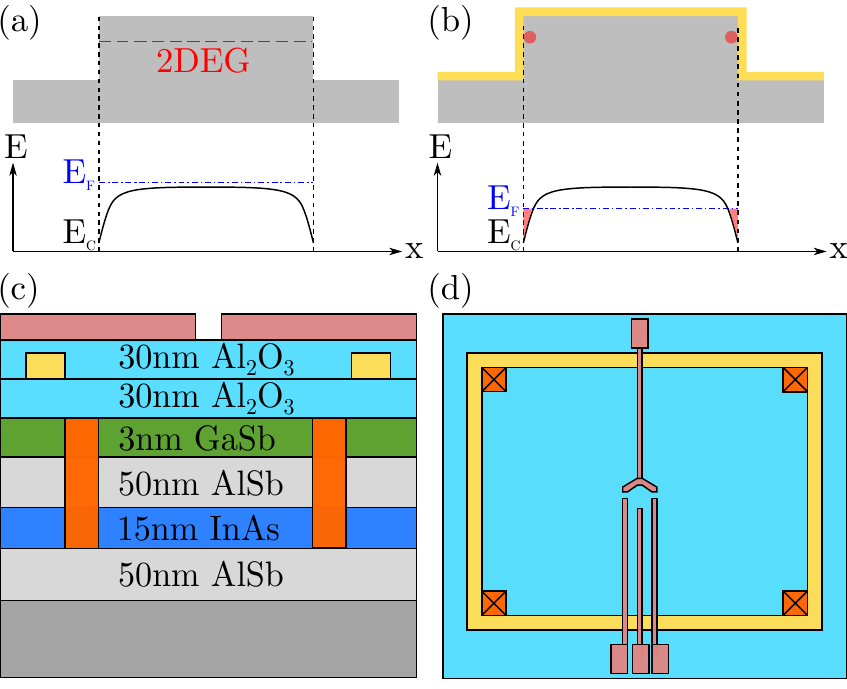}
	\caption{(a) InAs two-dimensional electron gas (red dashed line) confined by the edges of an etched mesa (top), and corresponding band edge diagram (bottom). (b) Same structures as in (a), but with a gate (yellow) on top charged with a negative voltage, thus depleting the two-dimensional electron gas. Due to the pinned Fermi level, electrons accumulate in the triangular potential well at the edges (shaded red in band edge diagram), leading to trivial edge states (red dots in top panel). (c) Schematic view of the cross-section of the heterostructure and the fabricated layer sequence. Ohmic contacts are colored in orange, the frame gate in yellow, and the fine gates in light red. (d) Schematic top view of the sample showing the lateral structure consisting of Ohmic contacts in an exemplary 4-point configuration, the rectangular frame gate, and fine gates defining a quantum dot. The color code is the same as in (c).} 
	\label{fig1}
\end{figure}

The effect of this Fermi-level pinning is illustrated in Figs.\,\ref{fig1}(a) and (b). The top panel of Fig.\,\ref{fig1}(a) shows an InAs two-dimensional electron gas with an etched mesa, the bottom panel its band edge diagram as a function of the real space coordinate $x$. The edges of the etched mesa form a surface, where the Fermi level is pinned in the conduction band. When the bulk of the two-dimensional electron gas is populated by carriers, this does not cause any problems. However, if the gate pushes the Fermi level into the band gap in the bulk [Fig.\,\ref{fig1}(b)], the Fermi level remains pinned in the conduction band at the edges. This causes electron accumulation\,\cite{noguchi_intrinsic_1991, olsson_charge_1996} in the triangular potential well created at the edges [indicated in red in Fig.\,\ref{fig1}(b)]. When measuring transport through such a structure (into the plane of Fig.\,\ref{fig1}), these accumulated electrons form a parasitic channel which conducts in parallel, as they are not or only weakly affected by the gate \cite{nichele_edge_2016, beukman_topology_2016}. The standard semiconductor nanofabrication approach is therefore not feasible in this case.

In this paper, we propose a gate geometry that circumvents the parasitic edge conduction.
We use a heterostructure containing a $\SI{15}{nm}$ InAs quantum well with an electron density $n=\SI{5.5e-11}{cm^{-2}}$ and a mobility of $\mu=\SI{1e5}{cm^2/Vs}$ in-between two \SI{50}{nm} AlSb barriers with a \SI{3}{nm} GaSb capping layer. The heterostructure composition below the lower barrier is identical to the one used in Ref.\,\onlinecite{mittag_passivation_2017}. The layer and fabrication sequence explained in the following is depicted in the schematic cross-section of the device shown in Fig.\,\ref{fig1}(c) and in the schematic top view displayed in Fig.\,\ref{fig1}(d). In a first step, Ohmic contacts of (Ge/Au/Ni/Au) were deposited. Then, a \SI{30}{nm} dielectric layer of $\mathrm{Al}_{\mathrm{2}}\mathrm{O}_{\mathrm{3}}$ was deposited by atomic layer deposition at a temperature of \SI{150}{\celsius}. In the next we deposited a frame-shaped gate of Ti/Au (\SI{10/70}{nm}) on the outside of the Ohmic contacts, which will be referred to as frame gate in the following. The frame gate is paramount to circumventing the trivial edge conduction. Upon depletion of the electron gas underneath, the inner part of the sample containing the Ohmic contacts without a physical edge or surface at which electrons could accumulate, is decoupled from the outside part.

\begin{figure}[]
	\includegraphics[width=\columnwidth]{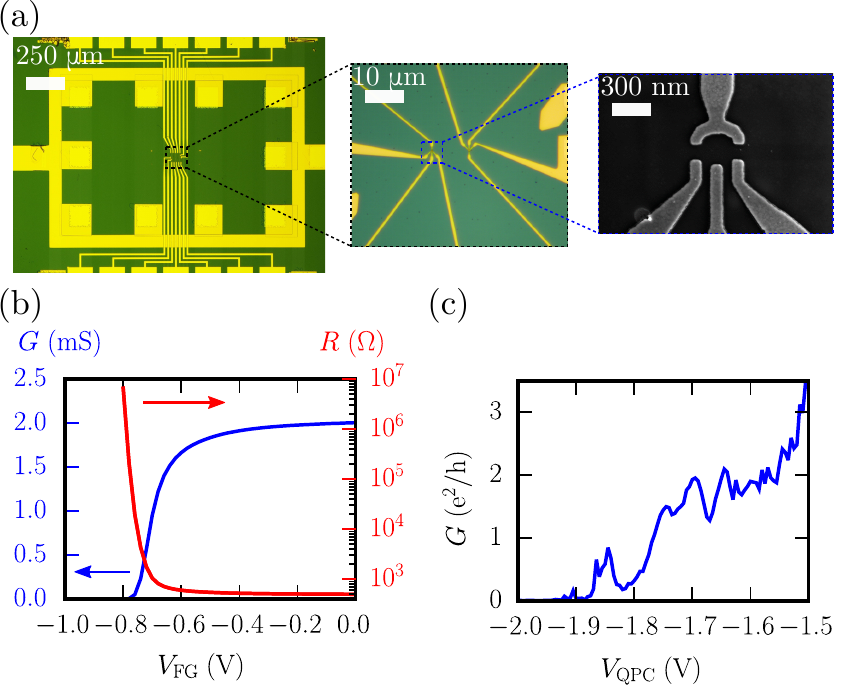}
	\caption{(a) Optical (left, center) and scanning electron micrograph (right) of a sample similar to the quantum dot sample used in this paper. (b) Conductance (blue) and resistance (red) between two contacts on the in- and outside of the frame gate as a function of voltage applied to the frame gate. (c) Conductance through a quantum point contact as a function of the voltage applied symmetrically to the split gates.}  
	\label{fig2}
\end{figure}

The deposition of a second \SI{30}{nm} $\mathrm{Al}_{\mathrm{2}}\mathrm{O}_{\mathrm{3}}$ film by atomic layer deposition allows for depositing fine Ti/Au (\SI{5/25}{nm}) gates forming a nanostructure on top [see Fig.\,\ref{fig2}(a)], employing electron beam lithography and metal evaporation. The nanostructures measured in this manuscript are a quantum point contact consisting of a split gate with \SI{200}{nm} separation and a quantum dot formed by a semicircular top gate and three finger gates forming a right and left barrier and a plunger gate. An optical and scanning electron micrograph of a finished sample are shown in Fig.\,\ref{fig2}\,(a).

We first characterize the frame gate in order to show that it completely disconnects the two-dimensional electron gas bulk from the boundaries. These measurements have been done at $T=1.3\,\mathrm{K}$ using AC lock-in techniques. We apply a bias voltage between a contact on the outside part of the two-dimensional electron gas [not shown in Fig.\,\ref{fig1}(d)] and the contacts on the inside. The results can be seen in Fig.\,\ref{fig2}\,(b) where we plot the conductance (blue curve and axis) and the resistance (red curve and axis) as a function of $V_{\mathrm{FG}}$, the voltage applied to the frame gate. At $V_{\mathrm{FG}}=0.8\,\mathrm{V}$, the conductance drops to zero as the two-dimensional electron gas underneath the frame gate is fully depleted. This voltage agrees with the expected depletion voltage taking into account the capacitance and electron density of the structure. At the same voltage, the resistance increases rapidly up to $10^7\,\Omega$, which was the maximum detectable resistance in our measurement setup. From this result we deduce that the regions of the two-dimensional electron gas in- and outside the frame gate are sufficiently decoupled from each other. Together with a full pinch-off in a quantum point contact measurement this will prove that our gate geometry circumvents the parasitic trivial edge conduction. 

In order to investigate whether this is the case, we measure a split gate quantum point contact in a 4-terminal geometry using AC lock-in techniques at $T=1.3\,\mathrm{K}$. In Fig.\,\ref{fig2}\,(c) we show the conductance $G$ as a function of the voltage $V_{\mathrm{QPC}}$ applied to both quantum point contact gates. Full pinch-off can be reached at $V_{\mathrm{QPC}}=-1.95\,\mathrm{V}$, which demonstrates that there is no parasitically conducting channel present underneath the gates. Quantized conductance steps or a $0.7$-anomaly are not visible, due to the resonances likely caused by disorder in the channel.

\begin{figure}[]
	\includegraphics[width=\columnwidth]{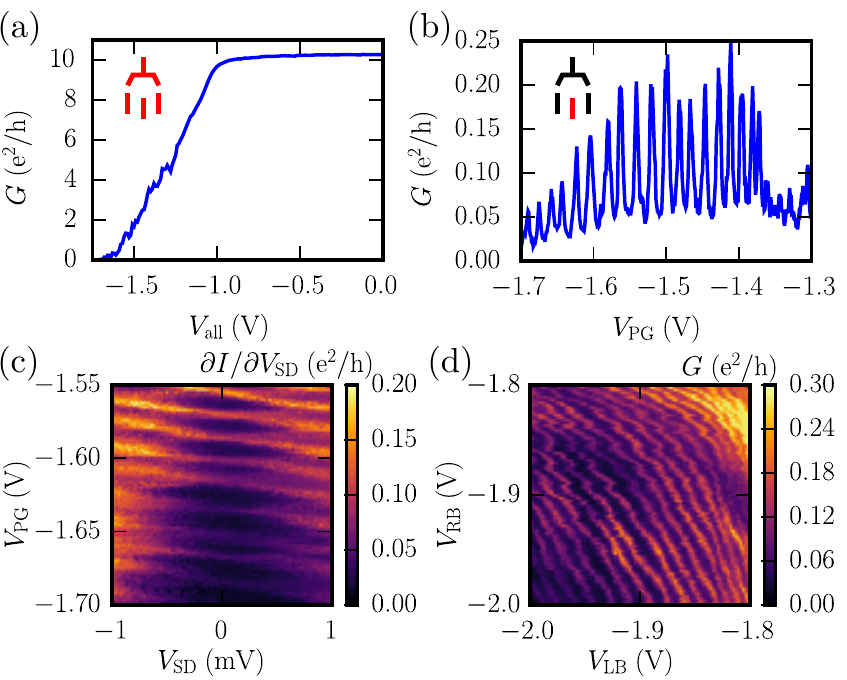}
	\caption{(a) Conductance $G$ through the quantum dot as a function of the voltage applied to all four quantum dot gates simultaneously (gates shaded in red in inset). (b) Coulomb resonances visible in the conductance $G$ through the quantum dot as a function of $V_{\mathrm{PG}}$ (c) Color plot of a finite bias measurement displaying the conductance $G$ through the quantum dot as a function of $V_{\mathrm{SD}}$ and $V_{\mathrm{PG}}$. (d) quantum dot tunability shown by changing both tunnel barriers in a color plot of the conductance $G$ as a function of $V_{\mathrm{LB}}$ and $V_{\mathrm{RB}}$.}  
	\label{fig3}
\end{figure}

In the following we investigate another sample with a frame gate. It contains finger gates for forming a quantum dot as seen in the scanning electron micrograph in the right panel of Fig.\,\ref{fig2}(a). These measurements were performed in a dilution refrigerator with a base temperature of $T=60\,\mathrm{mK}$ using AC lock-in techniques in a 2-terminal measurement. In a first step, we demonstrate that also the narrower quantum dot gates fully pinch off the two-dimensional electron gas. For this purpose, we apply a voltage $V_{\mathrm{all}}$ to all four quantum dot gates [colored red in the schematic of the gate layout in the inset of Fig.\,\ref{fig3}(a)] and operate the device as a quantum point contact. As seen in Fig.\,\ref{fig3}(a) the electron gas can be pinched off also in this geometry. Again resonances caused by disorder appear on the conductance curve.

We now tune the gates such that we form a small metallic island in-between the gates. We apply a voltage $V_{\mathrm{PG}}$ to the plunger gate to change the occupation of the quantum dot while measuring its conductance $G$. The result of this measurement is displayed in Fig.\,\ref{fig3}(b), where the red gate of the inset indicates the gate being swept. Sharp, evenly spaced conductance resonances indicate charging the quantum dot with single electrons.

In Fig.\,\ref{fig3}(c) we show a color plot of the quantum dot conductance while varying the applied DC source-drain bias voltage $V_{\mathrm{SD}}$ ($x$-axis) in addition to $V_{\mathrm{PG}}$ ($y$-axis). Coulomb-blockade diamonds are visible and from their extent in $V_{\mathrm{SD}}$ we can determine a charging energy $E_{\mathrm{C}} \approx 1\,\mathrm{meV}$ which agrees with estimations based on the capacitance of the island. This is an approximate value, as the outlines of the Coulomb diamonds are not very sharp and they increase in size for more negative $V_{\mathrm{PG}}$. From the size of the quantum dot and the electron density in the wafer we estimate the number of electrons in the dot to be $N \approx 400$. Neither excited states nor signatures of single particle levels are visible.

In Fig.\,\ref{fig3}(d), the conductance of the quantum dot is shown as a function of the voltages $V_{\mathrm{LB}}$ and $V_{\mathrm{RB}}$ applied to the left and the right barrier, respectively, in order to elucidate the tunability of the system. Charging both gates to more negative voltages we pass multiple Coulomb resonances, indicating that electrons are being expelled from the dot. The overall conductance decreases when both voltages are more negative, which is a sign of closing the tunnel barriers and therefore signals standard quantum dot operation.

Using the measurements Figures\,\ref{fig2}(c) and \ref{fig3}(d) we determine the lever arms of the quantum dot gates to be $\alpha_{\mathrm{PG}}=0.05$ for the plunger gate, $\alpha_{\mathrm{LB}}=0.08$ for the left barrier, and $\alpha_{\mathrm{RB}}=0.06$ for the right barrier, respectively. These values are within expectations considering the distance of the two-dimensional electron gas from the gates and the geometry of the structure. This indicates standard quantum dot operation in InAs quantum wells, like in technologically more mature two-dimensional electron gas systems like GaAs and Si. Such a performance has not been demonstrated before to the best of our knowledge.  

In conclusion, we demonstrated a technique which allows operating nanostructures such as quantum point contacts and quantum dots in InAs quantum wells by circumventing parasitic trivial edge conduction. This was achieved by separating the sample edge from the bulk of the two-dimensional electron gas with the frame gate.
Quantitative analysis of the nanostructure conductance is still severly limited by disorder. It may be overcome by improving material quality\,\cite{thomas_high_2018}, which is independent of the device geometry introduced here. The frame gate could in principle also be applied to other narrow-band material systems that suffer from similar undesired edge conduction or Fermi-level pinning issues.

%
   
\begin{acknowledgments}
The authors acknowledge the support of the ETH FIRST laboratory and the financial support of the Swiss Science Foundation (Schweizerischer Nationalfonds, NCCR QSIT).
\end{acknowledgments}

\bibliography{bibl}
\end{document}